\documentclass[agums,11pt]{aguplus}		

\usepackage{lineno}

\usepackage{amsmath,times,afterpage,color}
\usepackage[dvips]{graphicx}

\lefthead{CLAUDEPIERRE ET AL.}
\righthead{ULF WAVES DRIVEN BY VELOCITY SHEAR}

\received{}
\revised{}
\accepted{}

\journalid{}{}
\articleid{}{}
\paperid{}
\ccc{}
\cpright{}{}

\slugcomment{Submitted to {\it Journal of Geophysical Research}, 2007.}

\linenumbers*[1]

\begin{document}


\def\eq#1{Equation~(\ref{eq:#1})}
\def\fig#1{Figure~\ref{fig:#1}}
\def\tbl#1{Table~\ref{tbl:#1}}
\def\ch#1{Chapter~\ref{ch:#1}}
\def\sec#1{Section~\ref{sec:#1}}

\sectionnumbers

\title{Solar Wind Driving of Magnetospheric ULF Waves: Pulsations Driven by Velocity Shear at the Magnetopause}

\author{S. G. Claudepierre, \altaffilmark{1,2}
	S. R. Elkington, \altaffilmark{1}
	and M. Wiltberger \altaffilmark{3}}

\altaffiltext{1}{Laboratory for Atmospheric and Space Physics, 		University of Colorado, Boulder, CO USA.}
\altaffiltext{2}{Department of Applied Mathematics, 
	University of Colorado, Boulder, CO USA.}
\altaffiltext{3}{High Altitude Observatory, National Center for 	Atmospheric Research, Boulder, CO USA.}

\authoraddr{S. G. Claudepierre,
	Laboratory for Atmospheric and Space Physics,
	University of Colorado, 
	1234 Innovation Drive,
	Boulder, CO 80303;
	email: claudepi@colorado.edu}
\authoraddr{S. R. Elkington,
	Laboratory for Atmospheric and Space Physics,
	University of Colorado, 
	1234 Innovation Drive,
	Boulder, CO 80303;
	email: scot.elkington@lasp.colorado.edu}
\authoraddr{M. Wiltberger,
	National Center for Atmospheric Research,
	High Altitude Observatory,
	3450 Mitchel Lane,
	Boulder, CO 80301;
	email: wiltbemj@ucar.edu}

\begin{abstract}
  We present results from global, three-dimensional magnetohydrodynamic (MHD) simulations of the solar wind/magnetosphere interaction.  These MHD simulations are used to study ultra low frequency (ULF) pulsations in the Earth's magnetosphere driven by shear instabilities at the flanks of the magnetopause.  We drive the simulations with idealized, constant solar wind input parameters, ensuring that any discrete ULF pulsations generated in the simulation magnetosphere are not due to fluctuations in the solar wind.   The simulations presented in this study are driven by purely southward interplanetary magnetic field (IMF) conditions, changing only the solar wind driving velocity while holding all of the other solar wind input parameters constant.  We find surface waves near the dawn and dusk flank magnetopause and show that these waves are generated by the Kelvin-Helmholtz (KH) instability.  We also find that two KH modes are generated near the magnetopause boundary.  One mode, the magnetopause KH mode, propagates tailward along the magnetopause boundary.  The other mode, the inner KH mode, propagates tailward along the inner edge of the boundary layer (IEBL).  We find large vortical structures associated with the inner KH mode that are centered on the IEBL.  The phase velocities, wavelengths, and frequencies of the two KH modes are computed.  The KH waves are found to be fairly monochromatic with well defined wavelengths.  In addition, the inner and magnetopause KH modes are coupled and lead to a coupled oscillation of the low-latitude boundary layer.  The boundary layer thickness, $d$, is computed and we find maximum wave growth for $kd$ = 0.5--1.0, where $k$ is the wave number, consistent with the linear theory of the KH instability.  We comment briefly on the effectiveness of these KH waves in the energization and transport of radiation belt electrons.

\end{abstract}


\section{Introduction \label{sec:swdulf}}

One of the outstanding questions in the study of magnetospheric ultra low frequency (ULF) pulsations is the nature of their generation.   Throughout this paper, when we refer to ``ULF pulsations'' we are referring to any broadband or quasi-monochromatic pulsation in the range 0.5--15 mHz (Pc4--Pc5 bands, as defined by \citet{jacobs:64a}).  Several authors have shown that conditions in the solar wind are well correlated with ULF pulsations observed in the magnetosphere.   For example, \citet{mathie:01a} show a strong correlation between solar wind speed and ULF pulsation power in the dayside magnetosphere, for $L$ shells in the range $L$ $\approx$ 4--7.  The authors note that this high correlation is strong evidence that the Kelvin-Helmholtz (KH) instability at the magnetopause is the source of the pulsation energy.  \citet{kepko:03a} conducted a study of a series of events in which ULF pulsations were observed in the dayside magnetosphere at a discrete set of frequencies.  A spectral analysis of the solar wind density during the same time periods revealed significant wave power at the same set of discrete frequencies.  This relationship suggests that variations in the solar wind dynamic pressure are responsible for driving ULF pulsations in the dayside magnetosphere.  In addition, ULF variations in the Earth's convection electric field may respond directly to variations in the orientation and strength of the interplanetary magnetic field (IMF) \citep{ridley:97a,ridley:98a}. 

The suggested solar wind sources of magnetospheric ULF pulsations can be subdivided into three distinct driving mechanisms:  pulsations observed near the dawn and dusk flank magnetopause driven by the strong velocity shear present there; pulsations in the dayside, driven by variations in the solar wind dynamic pressure; and pulsations driven by variations in the orientation and strength of the IMF.  ULF pulsations generated by these different mechanisms are thought to occur primarily over different, but sometimes overlapping, local time sectors \citep{takahashi:92a,lessard:99b,ukhorskiy:05a}.  Thus, the global distribution of ULF wave power in the magnetosphere is an important diagnostic for understanding the generation mechanism(s).

The solar wind sources outlined above can be classified as external sources of ULF pulsations in the magnetosphere.  In addition to these proposed external sources, a number of authors have suggested that processes internal to the magnetosphere may also be responsible for the generation of magnetospheric ULF pulsations.  Wave particle interactions and local reconfigurations of the magnetic field are but two examples of a number of proposed internal sources, see the review by \citet{takahashi:98a} for more information.  The focus of this paper will be on external driving of magnetospheric ULF pulsations and internally generated pulsations will not be discussed further.

The spatial overlap of the distribution of ULF wave power for the different generation mechanisms complicates the study of the individual generation mechanisms.  For example, it could be argued that a satellite measurement of a ULF pulsation in the dayside, near the dusk flank, was generated by either an impulsive variation in the solar wind density or driven by velocity shear, through the KH instability.  Thus, a detailed knowledge of the upstream solar wind parameters is essential in determining the source of the ULF pulsation.   This highlights one of the main difficulties in studying the three generation mechanisms proposed: there are very few events in which one of the three solar wind generating parameters is dominant over the other two.  The solar wind is filled with complex structures and is quite dynamic.  Typically all three of the suggested mechanisms are operating simultaneously.

To circumvent these issues, we present results from a controlled experiment study of ULF pulsations in the magnetosphere.  We drive the Lyon-Fedder-Mobarry (LFM) global, three-dimensional, MHD simulation of the solar wind/\-mag\-neto\-sphere interaction with idealized solar wind conditions.  These idealized solar wind input parameters are chosen to mimic each of the three driving mechanisms outlined above.  By holding all of the solar wind input parameters constant except one, we are able to study the effect of changing only that one parameter.  The characteristics of the ULF pulsations generated by the particular driving mechanism under consideration can then be studied without the complications described above.  The focus of this paper will be on ULF pulsations driven by the strong velocity shear near the dawn and dusk flank magnetopause. 

Magnetospheric ULF pulsations are also known to be important in the energization and transport of radiation belt electrons.  \citet{rostoker:98a} showed a strong correlation between outer zone electron flux and magnetospheric wave power in the ULF band.  \citet{baker:98b} similarly noted an association between ULF wave power and energetic electron enhancements in a comparison of two magnetic cloud events.  For radiation belt electrons drifting in the equatorial plane, the most relevant field quantities for particle energization are the GSM z component of the magnetospheric magnetic field, $B_{z}$, and the GSM azimuthal component of the magnetospheric electric field, $E_{\phi}$ \citep{northrop:63a}.  Thus, our efforts to characterize the ULF pulsations generated in the LFM simulations will be focused on pulsations in these two magnetospheric field components.  Throughout this paper, we will comment on applications to radiation belt electron energization and transport, when appropriate.

The remainder of this paper is structured as follows:  In \sec{khi} we discuss the main theoretical and numerical work regarding the Kelvin-Helmholtz (KH) instability at the Earth's magnetopause.  \sec{lfm} provides a brief description of the global MHD simulations used in this study.  In \sec{simres} we present the simulation results along with the spectral analysis techniques that are used to study the ULF waves in the simulation magnetosphere.  \sec{discuss} compares the simulation results with the theoretical KH results from \sec{khi}.  In \sec{conclusions} we provide a brief summary and concluding remarks.

\section{The Kelvin-Helmholtz Instability at the Magnetopause \label{sec:khi}}

The Kelvin-Helmholtz instability occurs at the interface between two fluids in relative motion \citep{chandrasekhar:61a}.  \citet{dungey:55a} suggested that portions of the magnetopause boundary might be KH unstable. Observational evidence suggesting a KH-type interaction at the magnetopause boundary soon followed.  Surface waves \citep{aubry:71a,lepping:79a,fairfield:79a,sckopke:81a} and vortical structures \citep{hones:81a,saunders:83a} were observed propagating anti-sunward along the magnetopause boundary. 

Early theoretical attempts to describe the KH interaction at the magnetopause boundary were done by \citet{sen:63a}, \citet{fejer:64a}, and \citet{southwood:68a}.  These linear MHD treatments all assumed the boundary interface between the magnetospheric and magnetosheath plasmas to be a tangential discontinuity (TD).  A tangential discontinuity is a one dimensional layer with velocities everywhere parallel to the planar interface.  The total pressure and normal magnetic field are continuous across the interface.  All three studies attempted to quantify the effects of compressibility and found that for large relative flow velocities compressibility had a stabilizing effect.  This is analogous to hydrodynamic KH where it is well known that compressibility has a stabilizing effect \citep{chandrasekhar:61a}.  This early work resulted in a necessary condition for the onset of the KH instability at the magnetopause boundary, which is valid for incompressible plasmas separated by a tangential discontinuity \citep{hasegawa:75a}:

\begin{equation}
        ({\bf k} \cdot {\bf v})^{2} > \frac{1}{\mu_{o} m_{i}} \bigg(\frac{1}{n_{1}} + \frac{1}{n_{2}} \bigg) \bigg[({\bf k} \cdot {\bf B_{1}})^{2} + ({\bf k} \cdot {\bf B_{2}})^{2}\bigg]
\label{eq:hasegawa}
\end{equation}
{\bf B} is the magnetic field, $n$ is the number density, $\mu_{o}$ the permeability of free space, $m_i$ the ion mass, {\bf k} is the wave vector, and {\bf v} the relative velocity between the two plasmas ({\bf v}={\bf v}$_{1}$-{\bf v}$_{2}$).  In \eq{hasegawa}, the units are mks and the coordinate system is Cartesian with the boundary interface (e.g. the magnetopause) assumed to be planar.  We define the boundary interface to be the YZ plane where the Y axis lies in the GSM equatorial plane, parallel to the boundary (positive tailward), the Z direction parallel to the GSM z direction and the X direction normal to the planar interface.  Thus, the X and Y axes lie in the GSM equatorial plane, with the Y axis parallel to the boundary and the X axis normal to the boundary.  In \eq{hasegawa}, the subscripts 1 and 2 refer to the regions on either side of the planar interface, the YZ plane.  We define X $>$ 0 to be region 1 and X $<$ 0 to be region 2.  Along the dusk magnetopause, X $>$ 0 (region 1) corresponds to magnetosheath plasma and X $<$ 0 (region 2) corresponds to magnetospheric plasma.  The wave vector {\bf k} is restricted to the YZ plane (the boundary interface).    In what follows, we reserve capital XYZ for this boundary coordinate system and lowercase xyz for the standard GSM coordinate system used in our MHD simulations.  Strictly speaking, \eq{hasegawa} is only valid for incompressible plasmas separated by a tangential discontinuity;  however, many features of the KH instability are well approximated in this limit \citep{kivelson:84a}.
  
The early theoretical KH treatments of \citet{sen:63a}, \citet{fejer:64a}, and \citet{southwood:68a} all assumed a tangential discontinuity at the boundary interface.  However, satellite observations of magnetopause crossings revealed a thin, viscous boundary layer at the magnetopause, dubbed the low-latitude boundary layer \citep{hones:72a,akasofu:73a,eastman:76a}.  This boundary layer is roughly characterized by tailward flowing plasma on closed field lines.    The existence of a thin boundary layer near the magnetopause suggested that modeling the magnetopause as a tangential discontinuity was inaccurate.  In addition to this inaccuracy, a tangential discontinuity magnetopause cannot explain another key feature of observations:  monochromatic surface waves.  An incompressible KH model that assumes a TD at the boundary interface predicts a growth rate (\eq{hasegawa}, LHS-RHS) that is a monotonically increasing function of the wave number, $k$.  This implies a continuum of wavelengths will be excited and the smallest wavelength disturbances will grow the fastest.  This theoretical result contradicts magnetopause surface wave observations where monochromatic waves with well-defined wavelengths are typically seen \citep[e.g.][]{takahashi:91a,chen:93a}. 
 
The next level of sophistication in KH models came in the early 1980's where the effects of compressibility and/or a boundary layer of finite thickness were included.  The inclusion of either of these two effects complicated the calculations.  Either the calculation of the characteristic equation remained analytical but the roots, $\omega$ (the complex frequency), had to be solved for numerically \citep{lee:81a,pu:83a}. Or the linear MHD equations were reduced to an eigenvalue problem for $\omega$ and integrated numerically \citep{walker:81a,miura:82a}.

The KH theory of \citet{walker:81a} included a boundary layer of finite thickness and assumed compressible plasmas.  He showed that when the wavelength of the disturbance became comparable with the thickness of the boundary layer, the instability was quenched.  This implied a fastest growing mode at a particular value of $kd$, where $d$ is the boundary layer thickness.  He studied the interaction for several geometric configurations and reported the fastest growing mode occurred for $kd$ $\sim$ 1.   

Results from a similar study (boundary layer/\-com\-press\-ible plasmas) by \citet{miura:82a} found maximum wave growth for $kd$ $\approx$ 0.5--1.0 and were in good agreement with those of \citet{walker:81a}.  The reported values of $kd$ at which maximum wave growth occurs should be interpreted qualitatively when applied to the real magnetopause.  This is because the authors made various geometrical simplifications in their studies ({\bf B} $\mid \mid$ {\bf v}, {\bf B} $\perp$ {\bf v}) which are not always satisfied at the real magnetopause boundary.  However, the main result from these two studies is clear: the KH instability will become quenched when the wavelength of the disturbance becomes comparable with the boundary layer thickness, i.e. when $kd$ $\sim$ 1.  The value of $kd$ at which the instability becomes quenched corresponds to the value of $kd$ at which maximum wave growth will occur.  Note that this result implies a particular wavelength for the fastest growing mode and thus, a particular frequency for the fastest growing mode ($f$ = $v_{phase}$ / $\lambda$).  The inclusion of a boundary layer of finite thickness is thus able to explain the observations of monochromatic waves with a well-defined wavelength.  \citet{walker:81a} noted that the frequency of the fastest growing mode was in the Pc4--Pc5 range for typical values of $k$ and $d$, inferred from observations.

The inclusion of a boundary layer of finite thickness also allows for two KH modes to be generated at the boundary.  \citet{lee:81a} included a boundary layer of finite thickness in their study of incompressible KH at the magnetopause.  They reported that two KH modes were generated, one at the magnetopause boundary (the outer edge of the boundary layer) and one at the inner edge of the boundary layer (IEBL).  They referred to these two modes as the magnetopause mode and the inner mode, respectively.  They found the inner mode to be unstable most of the time whereas the excitation of the magnetopause mode depended critically on the orientation of the magnetic field in the magnetosheath.  It has been suggested that the vortical structures \citep{hones:81a} and the surface waves \citep{couzens:85a} observed near the magnetopause are associated with the KH instability at the IEBL.

\citet{pu:83a} gave a comprehensive study of compressible KH at the magnetopause boundary.  They assumed the boundary interface to be a tangential discontinuity and found two unstable KH modes, with different phase velocities and different wave vectors, {\bf k}.  They referred to these two modes as the fast and slow modes, where fast and slow refers to the different phase velocities.  As with previous authors, they found the addition of compressibility to have a stabilizing effect.  However, they found this effect to be small when compared with results in the incompressible limit (\eq{hasegawa}).  Their treatment also resolved the apparent discrepancies in the early work of \citet{sen:63a}, \citet{fejer:64a}, and \citet{southwood:68a} by recasting their results in terms of the slow and fast KH modes. 

A follow up paper \citep{kivelson:84a} discussed the results of \citet{pu:83a} in the context of \citet{lee:81a}.  They noted that when the magnetopause and IEBL were separated by a large distance (relative to the amplitude of the disturbance) the fast and slow modes of \citet{pu:83a} developed independently on the two interfaces.  However, when the magnetopause and the IEBL were close together, the fast and slow modes coupled giving rise to two new modes, one mode propagating on the magnetopause (magnetopause mode) and the other propagating on the IEBL (inner mode).  These two new modes had different phase velocities and different wavevectors, {\bf k}.  The phase velocity of the magnetopause mode was largely governed by the flow velocity in the magnetosheath while the inner mode phase velocity was governed by the flow velocity in the boundary layer. 

In what follows, we will compare results from global, three-dimensional MHD simulations of the solar wind/\-mag\-neto\-sphere interaction with the theoretical results detailed above.  We will demonstrate the existence of surface waves on the simulation magnetopause.  These surface waves will be shown to be driven by strong velocity shear and not dynamic pressure variations in the solar wind.  We will evaluate the condition for KH instability (\eq{hasegawa}) along the simulation magnetopause and show that it predicts the flow to be KH unstable at locations consistent with where the surface waves are seen in the simulation.  We will use spectral analysis techniques to compute the frequency of these surface waves.  We will also compute the wavelength of these surface waves directly from the simulation results.  A simulation boundary layer thickness will be computed and the results will be shown to be consistent with $kd$ = 0.5--1.0.  We will also show that two KH modes are excited near the simulation magnetopause boundary; one at the magnetopause and one at the inner edge of the boundary layer.  We will present a scientific visualization of the simulation results that shows both of these KH modes propagating tailward along their respective boundaries.  The scientific visualization will also reveal a coupled oscillation of the simulation boundary layer and large vortical structures associated with the inner KH mode.

\section{The LFM Global MHD Simulation \label{sec:lfm}}

The Lyon-Fedder-Mobarry (LFM) global, three-di\-men\-sion\-al magnetospheric model solves the single fluid, ideal magnetohydrodynamic (MHD) equations to simulate the interaction between the coupled magnetosphere - ionosphere system and the solar wind.  The details of the numerical methods used within the code are described in \citet{lyon:04a}.  As an inner boundary condition, the magnetospheric portion of the code couples to a 2D ionospheric simulation which computes the cross polar cap potential, needed for the plasma flow boundary condition, based upon the field aligned currents at the inner spherical boundary and empirical models for the extreme ultraviolet and auroral conductances.  The solar wind conditions, which form the outer boundary condition, can be taken from upstream satellite observations or can be created from scratch.  Runs with realistic solar wind inputs have been used to study geomagnetic storms \citep{goodrich:98a} and substorms \citep{lopez:98a}.  Idealized solar wind configurations have been particularly helpful in analyzing the physical processes involved in magnetospheric phenomenon, such as the erosion of the magnetopause \citep{wiltberger:03a}.

While the details of the numerical techniques used to solve the ideal MHD equations are beyond the scope of this paper, they do have an impact on the simulations ability to resolve boundary layers.  There are three key aspects of the numerical techniques used in the LFM that are important namely, the numerical order of the scheme, the use of nonlinear switches, and the size and shape of cells within the grid.   The numerical order of a scheme can be thought of as the accuracy of the interpolation in terms of a Taylor series.  A first order scheme introduces `numerical' diffusion into the solution, while higher order schemes avoid diffusion at the cost of dispersion errors which introduce artificial extrema into the solution.  Total variation diminishing (TVD) schemes are designed to balance the benefits of high and first order numerical schemes and are discussed in more detail in Chapter 21 of \citet{hirsch:88a}.  The LFM uses the Partial Donor Cell Method (PDM) \citep{hain:87a} as the nonlinear switch along with an eighth order interpolation scheme.  In a simple test with linear advection problems,  this approach allows for an increase by a factor of $400$ in the Reynolds number when compared with a simple first order scheme.  Since the numerical techniques used to solve the ideal MHD equations fall into the category of Finite Volume Methods, the cells used to discretize the computational domain are not required to be uniform or orthogonal.  This allows us to place regions of high resolution in areas known {\it a priori} to be important, e.g. the magnetopause.  In addition, these cells have aspect ratios designed to have more resolution in the directions transverse boundary than along it.  In practice the numerical order and use of the PDM switch in the LFM are not changed, but we can adjust the grid resolution.  In runs with the grid resolution changed by a factor of two in all directions we noticed roughly a 33$\%$  change in the thickness of the boundary layer.  Simulations with another factor of two increase in resolution are not practical at this time.

To investigate the ULF pulsations generated by the strong velocity shear at the dawn and dusk magnetopause, we drive the LFM simulation with a range of idealized solar wind input parameters.  The three LFM simulations used in this study differ only in the solar wind driving velocity.  The remaining solar wind driving parameters are identical for the three simulation runs: $B_{x}$ = $B_{y}$ = 0 nT, $B_{z}$ = -5 nT, $n$ = 5 particles/cm$^{3}$, $v_{y}$ = $v_{z}$ = 0 km/s, and sound speed = 40 km/s.  The three solar wind velocity inputs (corresponding to the three different simulations in this study) are $v_{x}$ = -400 km/s, $v_{x}$ = -600 km/s, and $v_{x}$ = -800 km/s.  These idealized solar wind conditions are chosen to represent moderate driving of the magnetosphere system under 3 different solar wind driving speeds.  In order to allow the magnetosphere to take shape within the simulation domain, the IMF $B_{z}$ component begins with an interval of southward IMF, turns northward, and remains southward for the remainder of the simulation interval.  The periods selected for analysis in this study are 4 hours long and occur two hours after the final southward turning of the IMF.  The solar wind input parameters listed above are held constant during the selected 4 hours.  Driving the simulations with constant solar wind parameters ensures that any discrete ULF pulsations in the simulation magnetosphere are not the result of perturbations in the solar wind.  In particular, the solar wind dynamic pressure is held constant in these three simulation runs.  Thus, any magnetopause surface waves that are generated cannot be the result of solar wind dynamic pressure fluctuations.  From here on, we will refer to the three different simulation runs as the 400, 600, and 800 runs.

The simulation results presented in this paper use a high resolution version of the magnetospheric grid.  While the spacing between cells is not uniform in the region near the magnetopause, the typical cell size is approximately 0.125 $R_E$ (Earth radii).  These simulations are conducted with idealized solar wind conditions with no dipole tilt in order to concentrate fully on the effects of velocity shear.  As has been described by \citet{korth:04a}, the LFM does not produce significant region 2 field aligned currents or a ring current, which means that the fields in the inner magnetospheric portion of the simulation will be more dipolar than is seen observations.   It also important to note that the LFM does not contain a plasmasphere and so the density profile in the inner magnetosphere will be different than the real magnetosphere.  While these differences are important, they will not prevent us from examining the structure and evolution of magnetospheric ULF oscillations at the magnetopause flanks in a realistic 3D configuration.

\section{Simulation Results \label{sec:simres}}

One of the advantages of this type of controlled parameter MHD study is the global, three-dimensional nature of the LFM MHD code.  Analyzing the results from the three simulations provides a global picture of the distribution of ULF pulsations in the inner magnetosphere, under the three different solar wind driving speeds.  We have developed a spectral analysis tool that provides a global map of where ULF pulsations occur in the simulation magnetosphere.  We briefly describe this tool and the spectral analysis techniques used therein.

\subsection{Spectral Analysis Techniques} \label{sec:specanal}

For the simulation field component of interest, say the simulation $B_{z}$, we record a 4 hour time series at every spatial point in the simulation domain.  At each spatial point, we compute the one-sided, periodogram power spectral density estimate, $P(f)$, of the zero-mean, 4 hour time series $x_{k}$, which we define as:

\begin{equation}
	P(f_{j}) = \frac{2\ dt}{N} \mid X_{j}\mid^{2} \quad for\ j = 0,1,\cdots,\frac{N}{2}
\label{eq:pow}
\end{equation}
where
\begin{equation*}
	X_{j}=\sum_{k=0}^{N-1}x_{k}exp[\frac{-2\pi ijk}{N}] \quad for\ j = 0,1,\cdots,N-1
\end{equation*}
and
\begin{equation*}
	f_{j}=\frac{j}{N\ dt} \quad for\ j = 0,1,\cdots,\frac{N}{2}
\end{equation*}
Here, $dt$ is the sampling rate in seconds, $f_{j}$ are the discrete Fourier frequencies in Hz, $N$ the number of points in the time series $x_{k}$, and $X_{j}$ the discrete Fourier transform (DFT) of the time series $x_{k}$.  If the units of $x_{k}$ are nT then the units of $P(f)$ are (nT)$^{2}$/Hz.  For the three LFM simulations in this study, these parameters are $dt$ = 30 seconds and $N$ = 480.  These sampling parameters determine the highest resolvable frequency, the Nyquist frequency, $f_{Ny}$ = 16.6667 mHz and the frequency resolution, ${\Delta}f$ = 0.0694 mHz.  

The result of this computation gives $P(f)$, the power spectral density estimate in the particular field component as a function of frequency, at every spatial point in the simulation domain.   We can now build a global picture of ULF wave power in a given frequency band by computing, at each spatial point, the integrated power ($IP$) over a given frequency band of interest $[f_{a}, f_{b}]$, via \eq{intpow}:

\begin{equation}
	IP = \int_{f_{a}}^{f_{b}} P(f) df
\label{eq:intpow}
\end{equation}
which has units (nT)$^{2}$ in this example.  Note that this quantity is different from the total power ($TP$) that is often used in ULF studies \citep[e.g.][]{engebretson:98a,mathie:01a}:

\begin{equation}
	TP = \sum_{j} P(f_{j}) \quad for\ all\ f_{j}\in [f_{a},f_{b}]
\label{eq:totpow}
\end{equation}
This quantity has units (nT)$^{2}$/Hz in this example and should more accurately be called a total power spectral density.  We favor \eq{intpow} over \eq{totpow} because \eq{totpow} does not explicitly account for the bandwidth, $df$.  A better definition of $TP$ would multiply the right hand side of \eq{totpow} by ($f_{b}-f_{a}$) and thus, would have units (nT)$^{2}$.  Finally, we note that Parseval's theorem can be expressed in this terminology as the root integrated power ($RIP$) of $P(f)$ equals root mean square ($RMS$) of the time series $x_{k}$:

\begin{align}
	\sqrt{\frac{1}{N}\sum_{k=0}^{N-1}x_{k}^{2}}&=\sqrt{\int_{0}^{f_{Ny}} P(f) df}\\\notag
(RMS &= RIP)
\label{eq:parseval}
\end{align}
where $f_{Ny} = 1/2dt$ is the Nyquist frequency.  

Computing power spectral densities from \eq{pow} often results in noisy spectra when plotted versus frequency.  Windowing the time series before computing the power spectral density estimate can smooth out this noisy behavior.  When we need to examine the finer frequency details of our power spectra, we first window the time series with the discrete prolate spheroidal sequences.  This spectral estimation method is commonly referred to as the `multi-taper method' \citep{thomson:82a,percival:93a}.

\subsection{Spatial Distribution of ULF Wave Power}

\begin{figure}[t!h!]
 \includegraphics[scale=2.0]{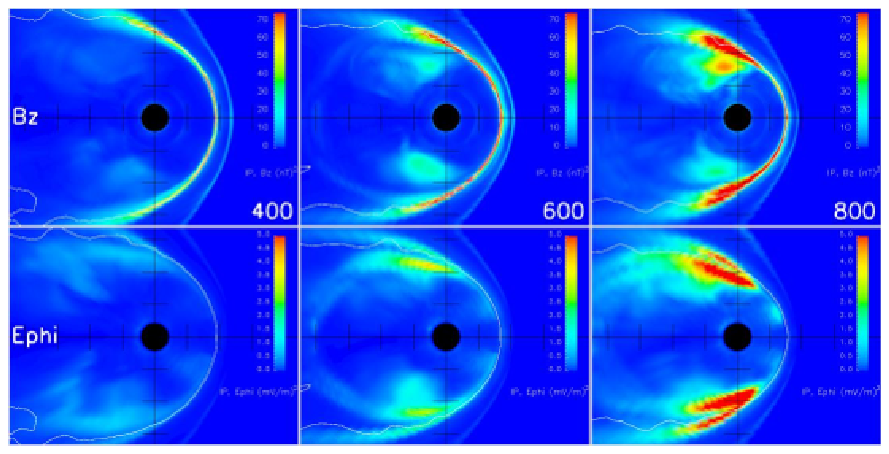}
 \caption{\label{fig:6panelpow} Global distribution of ULF integrated power ($IP$, 0.5--15 mHz)  in the GSM equatorial plane for the simulation $B_{z}$ (top row) and $E_{\phi}$ (bottom row) field components.  The three columns correspond to the three MHD simulations used in this study ($v_{sw}$ = 400 km/s, 600 km/s and 800 km/s, respectively).  The white contours are $B_{z}$=0 contours, the approximate location of the magnetopause.  The KH surface waves are manifest as the regions of intense $IP$ near the dawn and dusk flank magnetopause (sun to the right, 5 $R_{E}$ spaced ticks).  The $E_{\phi}$ $IP$ panels in the bottom row show the two distinct KH populations, the inner KH mode and magnetopause KH mode.  The color scales in each row are set to the same value to emphasize the increasing intensity of ULF wave power as the solar wind driving speed is increased.}
\end{figure}

\fig{6panelpow} shows the result of the prescribed technique for $B_{z}$ $IP$ (top row) and $E_{\phi}$ $IP$ (bottom row) ULF wave power, integrated over the frequency band 0.5 to 15 mHz (\eq{intpow}), for the 400, 600, and 800 km/s simulations (columns).  Each panel is a GSM equatorial plane cut with 5 $R_{E}$ spaced ticks on the x and y axes (sun to the right).  The black circle at the origin is the inner boundary of the simulation, located at r $\sim$ 2.2 $R_{E}$.  The $B_{z}$ $IP$ color scale ranges from 0 to 75 nT$^{2}$ and the $E_{\phi}$  $IP$ color scale ranges from 0 to 5 (mV/m)$^{2}$.  The color scales in each row are the same to emphasize the increasing intensity of ULF wave power as the solar wind driving velocity is increased.  The white contours in each of the panels in \fig{6panelpow} are $B_{z}$=0 contour snapshots, which for these idealized solar wind driving conditions, is the approximate location of the magnetopause:  The solar wind magnetic field is purely southward whereas the magnetospheric magnetic field is predominately northward.  Thus, the $B_{z}$=0 contour is good representation of the open/closed field line boundary.  The bow shock is also resolved as the region of ULF wave power upstream of the $B_{z}$=0 contour, particularly clear in the three $B_{z}$ $IP$ panels (top row).

\begin{figure*}[ht!]
 \includegraphics[scale=2.0]{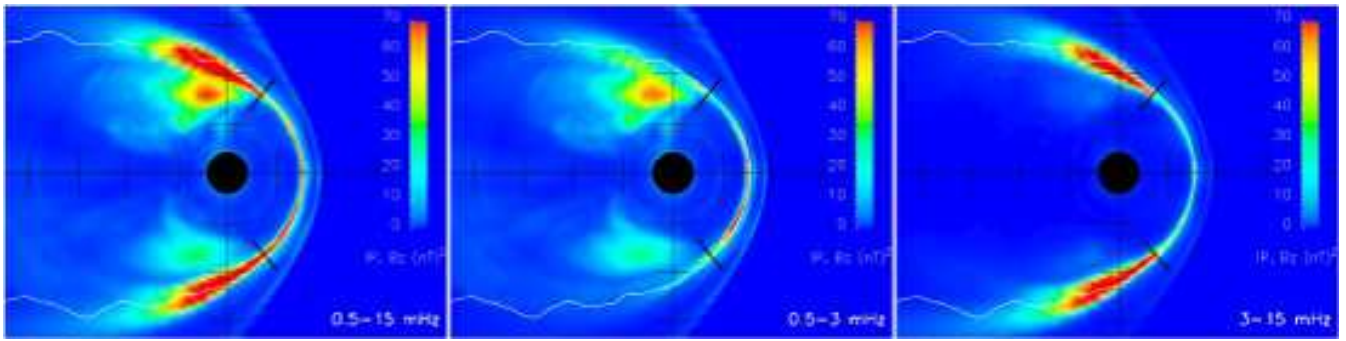}
  \caption{\label{fig:3panelpow} $B_{z}$ $IP$ in the GSM equatorial plane from the 800 km/s simulation, integrated over three different frequency bands to highlight the Kelvin-Helmholtz (KH) and magnetospheric (MSP) ULF wave populations.  The left panel is $B_{z}$ $IP$ integrated over 0.5--15 mHz (same panel as in \fig{6panelpow}).  The middle panel is $B_{z}$ $IP$ integrated over 0.5--3 mHz to highlight the MSP population.  The right panel is $B_{z}$ $IP$ integrated over 3--15 mHz to highlight the KH population.  In each panel, the two black lines perpendicular to the magnetopause mark the point along the magnetopause where the KH waves are first seen in the simulation.}
\end{figure*}

\fig{6panelpow} shows substantial ULF wave power in the $B_{z}$ and $E_{\phi}$ field components near the dawn and dusk flank magnetopause.  A close examination of the regions of intense ULF wave power shows that, in fact, there are three distinct ULF wave populations being driven in the simulations.  The first distinction can be seen in \fig{3panelpow}, which is taken from the 800 km/s simulation.  The leftmost panel in \fig{3panelpow} shows $B_{z}$ $IP$, integrated over the entire ULF band, 0.5--15 mHz (same panel as in \fig{6panelpow}).  The middle panel in \fig{3panelpow} shows $B_{z}$ $IP$, integrated over the frequency band 0.5--3 mHz.  The far right panel shows $B_{z}$ $IP$, integrated over 3--15 mHz. The higher frequency population  (3--15 mHz, \fig{3panelpow}, far right panel) is confined to the magnetopause boundary whereas the lower frequency population (0.5--3 mHz, \fig{3panelpow}, middle panel) is interior the magnetosphere, away from the boundary.  We will refer to the higher frequency (3--15 mHz) wave population generated near the magnetopause boundary as the Kelvin-Helmholtz (KH) population.  In \fig{3panelpow}, the two black ticks orthogonal to the magnetopause boundary mark the point on the magnetopause where the KH surface waves are first seen in the simulation.  The lower frequency population (0.5--3 mHz, \fig{3panelpow}, middle panel) is generated by a process internal to the magnetosphere.  This lower frequency ULF wave population along with its generation mechanism will be described in a follow up paper.  In what follows, we will refer to this lower frequency population as the magnetospheric (MSP) population.

The second distinction can be seen in the $E_{\phi}$ $IP$ panels (bottom row) in \fig{6panelpow} and is a distinction amongst the KH waves themselves.  A close examination of the KH population near the dusk flank magnetopause in the 800 km/s, $E_{\phi}$ panel (bottom right) in \fig{6panelpow} reveals two distinct wave populations being driven near the dusk flank magnetopause (also true at dawn).  In the panel, we see one region of intense ULF wave power aligned with the $B_{z}$=0 contour and a second, spatially larger region of ULF wave power earthward of the magnetopause boundary.  From here on, we will refer to the outer KH wave population, near the $B_{z}$=0 contour, as the magnetopause KH mode and the more earthward KH wave population as the inner KH mode.  We have verified that both the inner and magnetopause KH modes identified here in $E_{\phi}$ $IP$ are also identifiable in $v_{r}$ $IP$ (not shown), which ensures that there are indeed two distinct KH modes.

To summarize, we have identified 3 distinct ULF wave populations driven in the simulations: the MSP population (\fig{3panelpow}, middle panel) and the two KH modes (\fig{6panelpow}, bottom right), the magnetopause KH mode and the inner KH mode.  This three-fold distinction is true for all three simulations in this study (the 400, 600 and 800 runs).

\subsection{Spectral Distribution of ULF Wave Power}

\begin{figure*}[ht!]
  \includegraphics[scale=1.0]{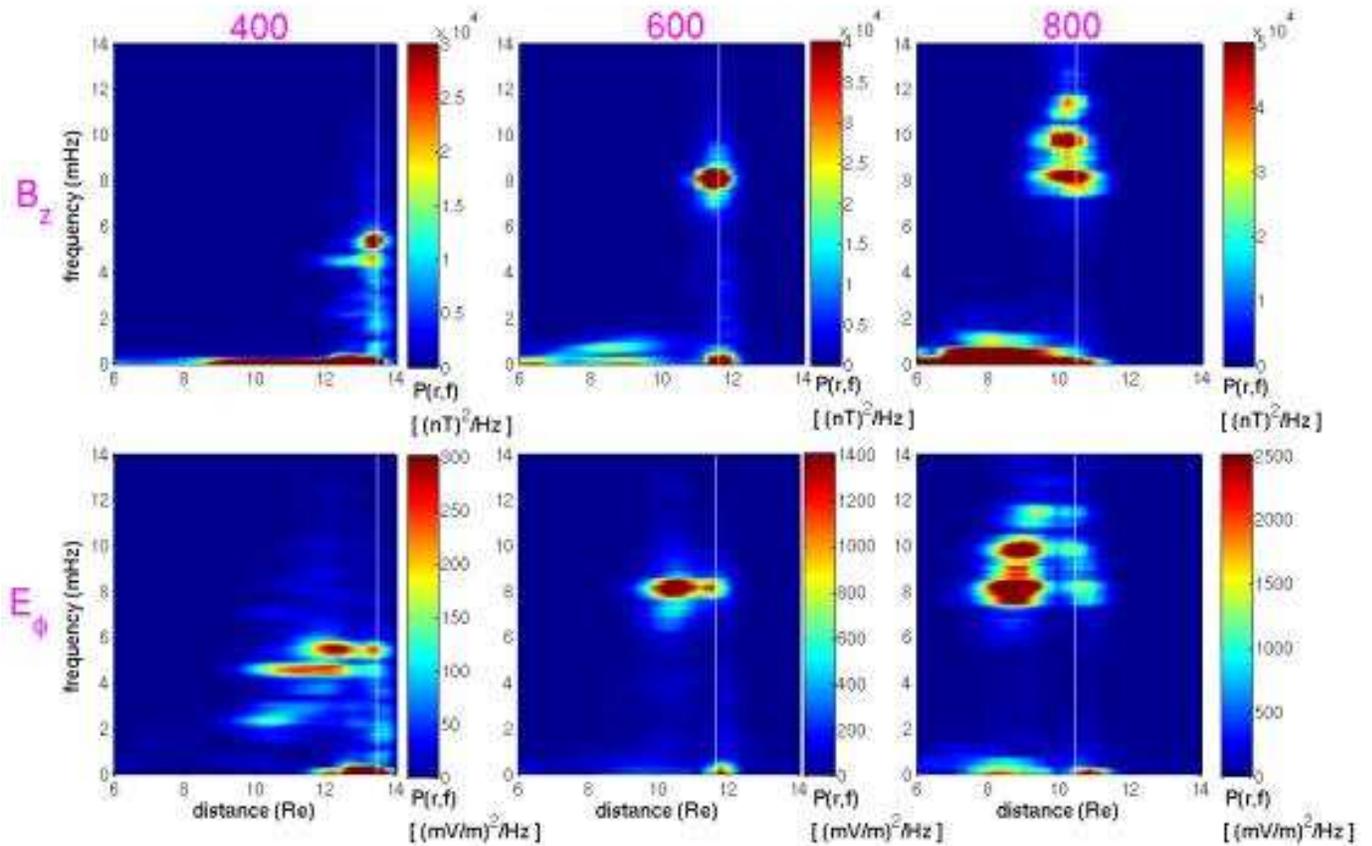}
  \caption{\label{fig:6panelfreq} Radial profiles of $B_{z}$ (top row) and $E_{\phi}$ (bottom row) power spectral density along the dusk meridian for the three simulations in this study (columns).  Distance along 18LT is on the horizontal axis, frequency is on the vertical axis and power spectral density is on the color scale.  The vertical white lines represent the approximate location of the magnetopause.  The KH waves excited near the magnetopause boundary are fairly monochromatic with peak frequencies near 5, 8, and 10 mHz for the 400, 600 and 800 km/s simulations, respectively.  The three $E_{\phi}$ panels in the bottom row show both the magnetopause KH mode (peaks in frequency near the magnetopause) and the inner KH mode (peaks in frequency earthward of the magnetopause).  Note the limited radial penetration depth of the inner KH mode (bottom row) and the uniform distribution of the MSP population (0.5--3 mHz) across a substantial portion of the dusk meridian (top row).}
\end{figure*}

We now describe the spectral distribution of the ULF waves in frequency and wave number space.  The spectral distribution of the ULF waves in frequency space is shown in \fig{6panelfreq}.  \fig{6panelfreq} shows radial profiles of $B_{z}$ (top row) and $E_{\phi}$ (bottom row) wave power spectral density (\eq{pow}) plotted along the dusk meridian for the 3 simulations (columns).  The horizontal axis is distance along 1800 local time (LT), the vertical axis is frequency from 0 to 14 mHz and wave power spectral density is plotted on the color scale.  The vertical white lines represent the approximate location of the magnetopause.  The ULF wave population excited near the magnetopause boundary is fairly monochromatic, with peak frequencies centered near 5, 8 and 10 mHz for the 400, 600 and 800 runs, respectively.  The color scales are all different in \fig{6panelfreq} so that the peak frequencies can be easily identified.  

The three $E_{\phi}$ panels in the bottom row of \fig{6panelfreq} show both the magnetopause KH mode and the inner KH mode described in the previous section.  The magnetopause KH mode is seen as the peaks in frequency centered on the white vertical lines (the approximate location of the magnetopause).  The inner KH mode is seen as the peaks in frequency earthward of the white vertical lines.  Note that the ULF wave power is more intense for the inner KH mode, which can also be seen in the bottom row of \fig{6panelpow}.  Also, the two KH modes have their peak power at the same frequencies, roughly  5, 8 and 10 mHz for the 400, 600 and 800 runs, respectively, which suggests that the two KH modes are coupled.

The three $E_{\phi}$ panels in \fig{6panelfreq} also show a limited radial penetration depth of the KH waves.  The inner KH mode penetrates roughly 3 $R_{E}$ inwards of the magnetopause boundary, for each of the three solar wind driving velocities.  This has implications for radiation belt transport and energization where equatorially drifting electrons can be energized by these ULF waves \citep{hudson:00a,elkington:03a}.  However, as \fig{6panelfreq} shows, this energization will only be effective within $\approx$ 3 $R_{E}$ of the magnetopause boundary.  This penetration depth is near the heart of the radiation belts (r $\approx$ 4--7 $R_{E}$) only for the 800 km/s simulation, where the magnetosphere is highly compressed.  However, the MSP population defined above (0.5--3 mHz, \fig{3panelpow}, middle plot) is distributed rather uniformly along the entire dusk meridian, particularly clear in the $B_{z}$ panels in the top row of \fig{6panelfreq}.  This population could effectively interact with radiation belt electrons through a drift resonant type interaction \citep{elkington:99a}.

An important quantity characterizing magnetospheric ULF pulsations is the azimuthal mode structure of the waves.  Determining the azimuthal mode structure up to mode number $m$ requires at least 2$m$ simultaneous satellite measurements, distributed in azimuth.  Thus, calculating the azimuthal mode structure from satellite measurements is especially difficult.  Global MHD simulations are not limited by these criteria and are well-suited to study the azimuthal mode structure over a large range of $m$ values.  To calculate the azimuthal mode structure, we follow the procedure outlined by \citep{holzworth:79a}.  This procedure is essentially a Fourier transform in space followed by a Fourier transform in time.  The spatial Fourier transform is done along circles of different radii.  The result of the full procedure gives $P(m,f)$, wave power spectral density as a function of azimuthal mode number and frequency, along different radii in the simulation domain.

\fig{3panelaz} shows the $E_{\phi}$ azimuthal mode structure for the 800 km/s simulation along three different radii in the simulation domain: 6.6, 8, and 10 $R_{E}$.  Here, and in \fig{6panelfreq} above, the multi-taper spectral estimate described in \sec{specanal} has been used.  In each of the color scale panels, the horizontal axis is azimuthal mode number from $m$ = 0--30, the vertical axis is frequency from 0--15 mHz and $E_{\phi}$ logarithmic power spectral density is on the color scale ( log($P(m,f)$) ).  The color scales are the same in each of the panels and range from -2 to 3.  The bottom panels beneath each of the color scale panels show integrated wave power over three different frequency bands, 0.5--3 mHz (green), 3--15 mHz (red), and 0.5 --15 mHz (blue), to distinguish between the MSP and KH populations.  The three panels in the figure show several interesting features.  First, along the radius of 6.6 $R_{E}$, we see the sub-3 mHz wave power, corresponding to the MSP population, and a hint of the KH populations near 10 mHz.  As we move further out in radius to 8 and 10 $R_{E}$, we begin to pick up the KH population near 10 mHz.  Second, the line plots underneath the three color plots show that the MSP and KH wave populations have their peak power at different azimuthal mode numbers.  The MSP population (0.5--3 mHz, green) typically has its peak wave power near $m$ $\approx$ 8 and does not extend much beyond $m$ $\approx$ 15.  On the other hand, the KH populations' (3 --15 mHz, red) wave power is distributed over a much broader range of $m$ values, say $m$ $\approx$ 0--30, with its peak near $m$ $\approx$ 15.  This feature is most evident in the 800 km/s, 10 $R_{E}$ panel (far right) where both populations are being sampled.  Similar features are seen in the $E_{\phi}$ azimuthal mode structure results from the 400 and 600 km/s simulations and in the $B_{z}$ azimuthal mode structure results from the 400, 600 and 800 km/s simulations (not shown here):  The MSP (0.5--3 mHz) population has its peak wave power near $m$ $\approx$ 8 and the KH populations' (3 --15 mHz) peak wave power is near $m$ $\approx$ 15.

It is interesting to note that the $m$ number of the peak wave power for both the KH and MSP populations does not vary significantly as the solar wind driving speed is varied.  This also has implications for radiation belt transport and energization where discrete peaks at a particular frequency and a particular mode number will select the particles that will be energized \citep{hudson:00a,elkington:99a}.  In particular, a given \{$m$,$f$\} pair will determine the drift frequency of the electrons that the KH waves could interact with, through the drift resonance condition: $\omega$ = $m$$\omega_{d}$.  Using this drift frequency, we can compute the relativistic first adiabatic invariant, $M$, for the given \{$m$,$f$\} pair.  This value of $M$ defines the particle population that could be energized by the KH waves.  For this calculation, we use the dipole approximation for the $L$ value and assume the electrons interact with the KH waves at the dusk meridian.  \tbl{innerMvals} and \tbl{outerMvals} show the results of this calculation at two different points along the dusk meridian. \tbl{innerMvals} shows the results for the most inward radial penetration of the inner KH mode.  \tbl{outerMvals} shows the results of the calculation for the inner KH mode near the magnetopause.  The values of the magnetic field, $B$, that are used in computing the relativistic correction factor are also shown.  A 1 MeV electron drifting in the equatorial plane near geosynchronous orbit has an $M$ value of roughly 1800 MeV/G.  The values of $M$ listed in \tbl{innerMvals} and \tbl{outerMvals} range from roughly 1/5 to 1/2 of this value.  We thus conclude that the KH waves could interact with equatorial plane electrons of a few hundred keV, near the $L$ values listed in the table.

\section{Discussion \label{sec:discuss}}

\subsection{Inner and Magnetopause KH Modes \label{sec:2khmodes}}

\begin{figure*}[ht!]
  \includegraphics[scale=2.0]{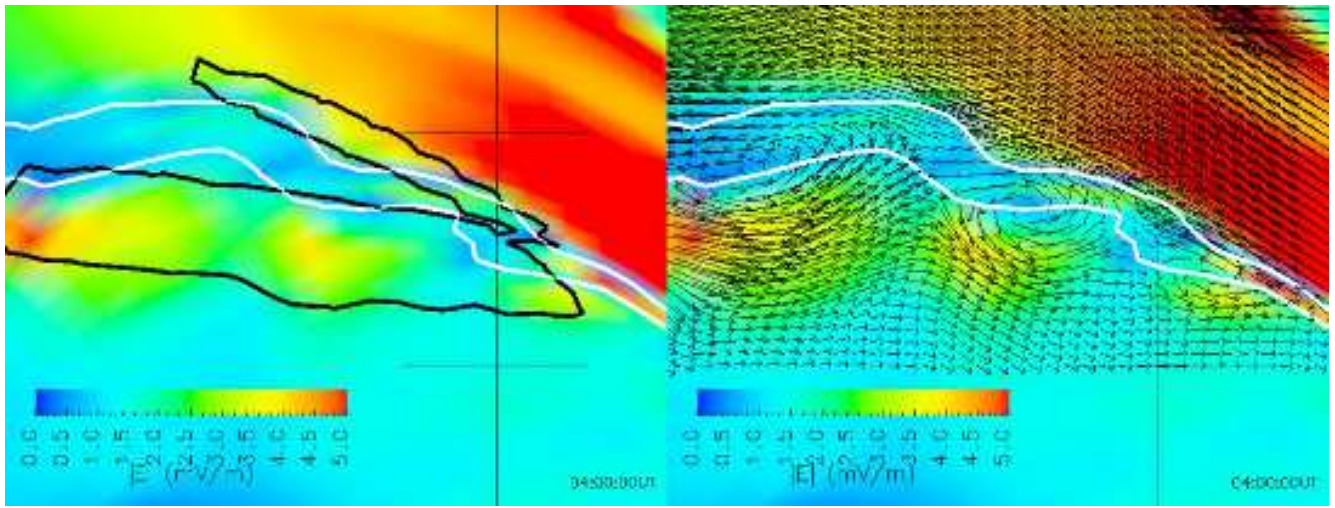}
  \caption{\label{fig:animsnap} Scientific visualization snapshot of the dusk flank magnetopause from the 400 km/s simulation.  The black vertical axis is the GSM positive y axis with ticks at 10 and 15 $R_{E}$ (sun to the right).  In both panels, {\textbar \bf E\textbar} is on the color scale from 0 to 5 mV/m.  The upper white contour is a $B_{z}$=0 contour, the approximate location of the magnetopause.  The lower white contour is a $v_{x}$=0 contour, the approximate location of the IEBL.  The region in between these two contours is the simulation boundary layer.  In the left panel, the two black $E_{\phi}$ $IP$ contours are shown to outline the inner and magnetopause KH modes.  The inner KH mode propagates tailward inside the larger black $E_{\phi}$ $IP$ contour, near the IEBL.  The magnetopause KH mode propagates tailward inside the smaller black $E_{\phi}$ $IP$ contour, near the magnetopause.  In the right panel, the $E_{\phi}$ $IP$ contours are replaced with the local velocity field.  Note that the counterclockwise oriented vortices are associated with the inner KH mode and centered on the IEBL.  These vortices grow in size as they propagate tailward from roughly 1.7 $R_{E}$ by 1.0 $R_{E}$ near the right side of the panel to roughly 5 $R_{E}$ by 3 $R_{E}$ near the left side of the panel.  Also note that the boundary layer thickens through the KH region, from roughly 0.5 $R_{E}$ near the right side of the panel to roughly 1.3 $R_{E}$ near the left side of the panel.}
\end{figure*}

In order to fully characterize the distinction between the two KH modes alluded to above, we must first define a magnetopause boundary layer in the simulation.  We define the simulation boundary layer (BL) as the continuous region of space that is 1. earthward of the $B_{z}$=0 contour (the magnetopause) and 2. where the local plasma flow is in the same sense as the local magnetosheath flow (tailward).  \fig{animsnap} is a GSM equatorial plane snapshot of the dusk flank magnetopause taken from the 400 km/s simulation (a scientific visualization of the simulation results can be downloaded here:   ).  This scientific visualization was created with the CISM-DX visualization package for OpenDX \citep{wiltberger:05a}.  The total electric field, {\textbar \bf E\textbar}, is on the color scale, ranging from 0 to 5 mV/m.  We choose to plot {\textbar \bf E\textbar} as opposed to $E_{\phi}$ because the two KH modes are most easily identified in {\textbar \bf E\textbar}.  The black vertical axis is the GSM positive y-axis, with ticks at 10 and 15 $R_{E}$ from bottom to top (sun to the right).  The upper white contour is the $B_{z}$=0 contour which is a very good approximation of the magnetopause in these idealized simulations.  The lower white contour is a $v_{x}$=0 contour.  Near the dusk flank this contour tracks the approximate delineation between tailward flowing (boundary layer) plasma and non-tailward flowing (magnetospheric) plasma.  The region between these two contours is approximately the simulation BL defined above.  The black contours are $E_{\phi}$ $IP$ contours, outlining the inner KH mode and magnetopause KH mode populations (\fig{6panelpow}, bottom row).  The two panels in the figure are identical except in the right panel we have replaced the $E_{\phi}$ $IP$ contours with the local velocity field.

The scientific visualization reveals the inner mode and magnetopause mode distinction described above.  We see both the inner and magnetopause KH modes propagating along their respective boundaries.    The inner mode is clearly seen as the tailward propagating blobs of {\textbar \bf E\textbar} inside the larger black $E_{\phi}$ $IP$ contour, just below the $v_{x}$=0 contour.  The magnetopause mode is less apparent.  It propagates tailward inside the smaller black $E_{\phi}$ $IP$ contour, just above the $B_{z}$=0 contour.  The coupled oscillation of the simulation boundary layer is striking.  The structure of the simulation boundary layer is very similar to the diagram of Model B presented in \citet{sckopke:81a}.  \citet{sckopke:81a} proposed 3 models (A, B, and C) of the low latitude boundary layer to explain ISEE observations.  Model A has both the magnetopause and the IEBL stable, Model B has both the magnetopause and the IEBL disturbed by surface waves and Model C has the magnetopause stable and the IEBL unstable.  Our scientific visualization clearly shows both the magnetopause and the IEBL to be disturbed by surface waves and the BL configuration thus corresponds to Model B.  A thickening of the simulation boundary layer through the KH region is also seen.  The simulation boundary layer thickness near the right side of each panel in \fig{animsnap} is roughly 0.5 $R_{E}$ and grows to roughly 1.3 $R_{E}$ near the left side of the panel.  

As discussed in the introduction to the \citet{safrankova:07a} paper on variations in boundary layer thickness, there are many open questions regarding the formation and structure of the low-latitude boundary layer.  \citet{song:92a} developed an explanation for the formation of the LLBL during strongly northward IMF that relies on magnetic reconnection at high latitudes. \citet{luhmann:84a} presented a discussion for the formation of the boundary layer on open field lines during southward IMF.  Older work of \citet{eastman:79a} indicated a role for viscous and diffusive mixing plasma from the magnetosheath onto closed field lines.  In our results we are seeing antisunward flow on closed field lines during a prolonged interval of southward IMF,  which we believe is reflection of the numerical viscosity.

The right panel in \fig{animsnap} (and in the scientific visualization) shows the counterclockwise oriented vortices propagating tailward in the simulation BL.  The orientation of the vortices is consistent with what is predicted by KH theory and with what has been observed near the magnetopause \citep{hones:78a,saunders:83a}.   Note that the vortices are associated with the inner KH mode and are centered on the $v_{x}$=0 contour, which is approximately the IEBL.  This fact has been alluded to many times \citep[e.g.][]{hones:81a,couzens:85a}.  Near the right side of each panel, where the KH waves are first seen in the simulation, a typical vortex size is roughly 1.7 $R_{E}$ in extent along the IEBL by roughly 1.0 $R_{E}$ in extent perpendicular to the IEBL.  The vortices grow in size as they move downtail and can grow to be as large as roughly 5 $R_{E}$ by 3 $R_{E}$ near the left side of the panels.  The ratio of the vortex dimensions in the equatorial plane remains constant at roughly 1.7 throughout the KH region.  

Kelvin-Helmholtz vortices are thought to be important for mass and momentum transport across the magnetopause, into the magnetosphere.  This proposed mechanism is particularly important for northward IMF conditions when reconnection is less effective in plasma transport across the boundary \citep{nykyri:01a,hasegawa:04a}.  It is certainly possible that the large vortical structures straddling the IEBL in the simulations could transport plasma into the magnetosphere.  However, in the present study, we make no attempt to quantify this possible  transport mechanism.

\begin{figure*}[ht!]
  \includegraphics[scale=1.0]{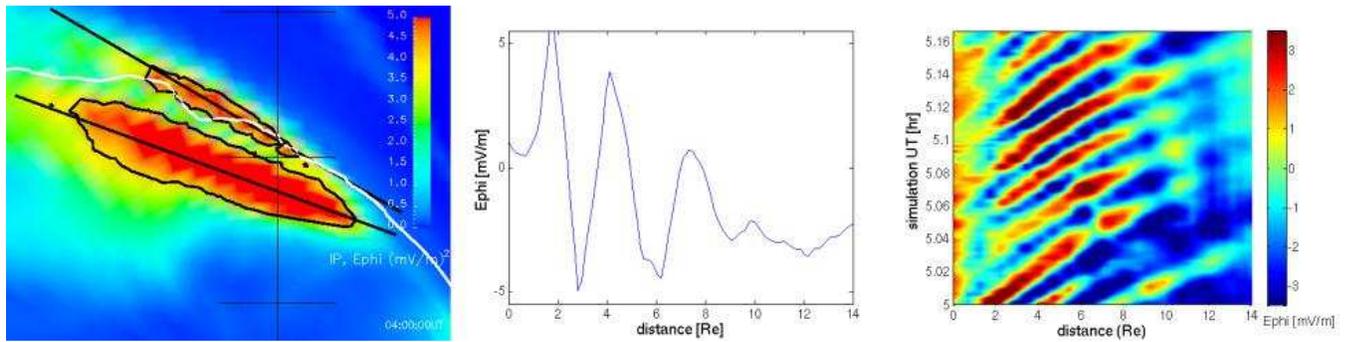}
  \caption{\label{fig:phasespeed} Example of how the wavelength and phase velocities are computed for the two KH modes in the simulations.  The left panel is a dusk flank zoom in of the $E_{\phi}$ $IP$ panel in \fig{6panelpow} for the 800 km/s simulation.  The two black lines in this panel define to the Y directions for the two KH modes.  The middle panel shows $E_{\phi}$ plotted along the black line for the inner KH mode, from which we calculate a wavelength of $\lambda_{Y}$ $\approx$ 3.3 $R_{E}$.  The right panel is a time series of plots shown in the middle panel.  The horizontal axis is distance along the inner mode black line (left panel), the vertical axis is simulation time and $E_{\phi}$ is on the colorscale.  By measuring the slope of the linear features in this panel, we calculate a phase velocity of $v_{phase}$ $\approx$ 225 km/s.  Note the coherent structure of the waves as they propagate downtail.}
\end{figure*}

The coupled oscillation of the two KH modes is strong evidence that these are in fact the two KH modes described in \citet{lee:81a} and \citet{kivelson:84a}.  In order to confirm this, we must show the two modes have different phase velocities and different wave vectors, {\bf k}.  We can extract the phase velocity and wavelength characteristics of the two KH modes directly from the simulation results.  By placing a line grid in the equatorial plane, through the two regions of $E_{\phi}$ $IP$ (\fig{phasespeed}, left panel) and plotting $E_{\phi}$ along this line, we can calculate the wavelength for each of the modes.  This corresponds to the wavelength in the Y direction in the boundary coordinate system defined in \sec{khi}.  The middle panel in \fig{phasespeed} shows this result for the inner KH mode in the 800 km/s simulation, from which we calculate a wavelength of $\lambda_{Y}$ $\approx$ 3.3 $R_{E}$.  The right panel in \fig{phasespeed} is essentially a time series of plots shown in the middle panel.  Distance along the equatorial line grid is plotted on the horizontal axis, simulation time along the vertical axis and $E_{\phi}$ is on the color scale.  By measuring the slope of the linear features in the plot, we calculate a phase speed of $\approx$ 225 km/s.  This panel also shows the coherent structure of the waves as they propagate downtail.  Using $\lambda_{Y}$ = 3.3 $R_{E}$ and $v_{phase}$ = 225 km/s, we calculate a wave frequency of $f$ $\approx$ $v_{phase}$/$\lambda_{Y}$ $\approx$ 11 mHz.  This calculation of the wave frequency is in good agreement with the peak frequency observed in the far right panels in \fig{6panelfreq}.  A similar calculation is done for the magnetopause mode in the 800 km/s simulation and for the inner and magnetopause modes in the 400 km/s and 600 km/s simulations.  The results are shown in \tbl{innerfreq} and \tbl{mpfreq}.  Note that the Y direction is slightly different for the two KH modes.  This is because the Y axis for each mode is chosen so that it is parallel to the boundary for that mode; for the magnetopause KH mode, this boundary is the magnetopause and for the inner KH mode, this boundary is the IEBL.  Thus, $\lambda_{Y}$ should be interpreted as wavelength along each respective boundary.  This slight difference can be seen in the left panel in \fig{phasespeed} as the two black lines (the respective Y axes) do not point in the same direction.

The frequencies listed in \tbl{innerfreq} and \tbl{mpfreq} are in good agreement with the peak frequencies in \fig{6panelfreq}.  This confirms that the two KH modes seen in the scientific visualization correspond to the two regions of $E_{\phi}$ ULF wave power near the dusk flank magnetopause in the bottom row of \fig{6panelpow}.  Moreover, when considering a particular simulation, the results in \tbl{innerfreq} and \tbl{mpfreq} show that the two KH modes have different phase velocities and different wavelengths but similar frequencies.  For example, in the 800 km/s simulation, we see that the phase velocity and wavelengths between the two KH modes differ by about 60 percent.  However, the difference between the two frequencies is only about 5 percent.  A similar result holds for the 400 km/s and 600 km/s simulations.  The coupled oscillation of the two KH modes is clear and we can positively identify the two surface modes in the simulation as the inner and magnetopause KH modes described in \citet{lee:81a} and \citet{kivelson:84a}.

\subsection{Boundary Layer Effects: Fastest Growing Mode \label{sec:bleffects}}

The results from the previous section, along with the direct power spectral density computations (\fig{6panelfreq}) show the two KH modes to be fairly monochromatic, with well-defined peak frequencies (\tbl{innerfreq} and \tbl{mpfreq}/\fig{6panelfreq}).  The monochromatic nature of the waves is a direct result of the presence of a boundary layer, as discussed in \sec{khi}.  Recall that the KH instability is quenched when $kd$ $\sim$ 1 where $k$ is the wave number and $d$ is the boundary layer thickness.  Thus, there is a particular wavelength (and frequency) for the fastest growing mode.  We can therefore explain our discrete KH frequencies ($\approx$5, 8, and 10 mHz for the 400, 600, and 800 km/s simulations, respectively) by showing that $kd$ $\sim$ 1 in our simulations.

We begin by defining the wave appearance region (WAR) of the KH waves as the point along the magnetopause where the KH waves are first seen in the simulations.  We determine these locations through a careful inspection of scientific visualizations from the three simulations.  These points are located at 1624 LT along the magnetopause for the 800 km/s simulation, at 1648 LT along the magnetopause for the 600 km/s simulation, and at 1708 LT along the magnetopause for the 400 km/s simulation.  For the 800 km/s simulation, this location is marked in the far right panel in \fig{3panelpow} with a black line perpendicular to the magnetopause.  We use our simulation results near these points to calculate the simulation boundary layer thickness, as defined in \sec{2khmodes}, at the WAR.

\begin{figure}[hr!]
  \includegraphics[scale=1.0]{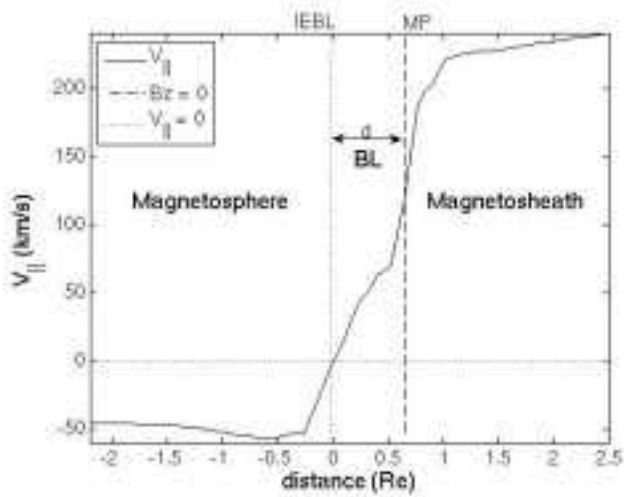}
  \caption{\label{fig:veloprof} Parallel velocity profile near the KH wave appearance region (WAR) for a particular timestep in the 400 km/s simulation.  The velocity parallel to both the magnetopause boundary and to the magnetosheath flow ($v_{\mid \mid}$) is plotted along a line perpendicular to the boundary (horizontal axis).  The vertical dotted line is the location of the IEBL while the vertical dashed line is the location of the magnetopause.  The region between these two lines is the simulation boundary layer, as defined in the text.  We see a boundary layer thickness, $d$, of roughly 0.65 $R_{E}$.}
\end{figure}

We compute the simulation boundary layer thickness near the WAR as follows:  At the WAR, we extract the local velocity profile along a line perpendicular to the boundary (for example, \fig{3panelpow}).  From this information, we compute the velocity locally parallel to both the magnetopause boundary and to the magnetosheath flow, in the equatorial plane.  \fig{veloprof} shows an example of this profile perpendicular to the magnetopause, at the WAR (1708 LT along the magnetopause), for a particular timestep in the 400 km/s simulation.  The solid line is the parallel velocity plotted against distance orthogonal to the boundary.  The vertical dashed line indicates the point on the line orthogonal to the boundary where $B_{z}$=0.  This is the location of the magnetopause for this particular timestep.  The vertical dotted line indicates the point on the line orthogonal to the boundary where the parallel velocity transitions from negative to positive values.   This is the location of the IEBL for this particular timestep.  The distance between these two vertical lines is the simulation boundary layer thickness, $d$, at the WAR.  In \tbl{innerkays} and \tbl{mpkays}, we show the results of this computation for the boundary layer thickness, $d$, at the dusk WAR, for the three simulations in this study.  We note that near the WAR, the simulation boundary layer thickness fluctuates throughout the 4 hour interval.  The values of $d$ listed in \tbl{innerkays} and \tbl{mpkays} are the average values for the 4 hours of simulation time and are typical values for the thickness depth.  It is also important to note that the LFM grid resolution near these points is sufficient to resolve this boundary layer thickness.  There are typically 3--4 grid cells within the simulation boundary layer.

\begin{figure}[r!b!]
  \includegraphics[scale=2.0]{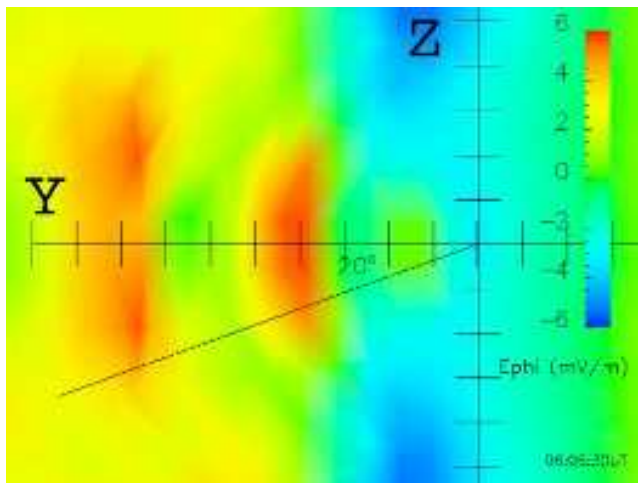}
  \caption{\label{fig:kz} The extent of the KH waves out of the equatorial plane, for the inner KH mode in the 800 km/s simulation.  The origin of the coordinate system is located at the KH wave appearance region (WAR), 1624 LT along the magnetopause.  The Y axis lies along the IEBL in the equatorial plane with the positive direction tailward.  The Z axis is parallel to the GSM z axis.  $E_{\phi}$ is on the color scale from -6 to 6 mV/m.  The axes ticks are spaced at 1 $R_{E}$.  Note that the KH waves are generated near the equatorial plane and propagate in both the Y and Z directions.}
\end{figure}

In order to evaluate $kd$, we must also calculate $k$.  In \eq{hasegawa}, the wave vector {\bf k} is restricted to the YZ plane.  We can approximate $k$ from our computed values of $\lambda_{Y}$ listed in \tbl{innerfreq} and \tbl{mpfreq} (i.e. $k$ $\approx$ $k_{Y}$).  This is a reasonable approximation, as can be seen in \fig{kz} where $E_{\phi}$ is plotted on the colorscale from -6 to 6 mV/m in the YZ plane for the inner KH mode in the 800 km/s simulation.  As described in \sec{khi}, the Y axis lies in the equatorial plane and is parallel to the boundary, in this case the IEBL.  The Z axis is parallel to the GSM z axis.  The origin of the coordinate system in \fig{kz} is located on the magnetopause at the WAR (1624 LT).  The axes ticks are spaced at 1 $R_{E}$ and a black line that makes a 20$^{\circ}$ angle with the equatorial plane is also shown.  Note that the KH waves are generated near the equatorial plane, which can be inferred through a careful inspection of \eq{hasegawa}.  Clearly, the KH waves propagate not only in the positive Y direction (tailward) but also in the Z direction.  This indicates a small $k_{Z}$ component to {\bf k}, in addition to $k_{Y}$ = 2$\pi$ /  $\lambda_{Y}$ and that the approximation $k$ $\approx$ $k_{Y}$ is valid.  The results of the $kd$ calculation are shown in \tbl{innerkays} and \tbl{mpkays}, under the assumption $k$ $\approx$ $k_{Y}$.  Our values of $kd$ are consistent with $kd$ $\sim$ 1.  This explains why we see KH waves of a particular wavelength (or frequency) in the simulation results.  The presence of the boundary layer of finite thickness quenches the KH instability and thus we have maximum growth for a particular $k$ and a particular frequency, $f$.  The monochromatic KH waves seen in the simulations are manifestations of this process.

\begin{figure*}[ht!]
  \includegraphics[scale=1.0]{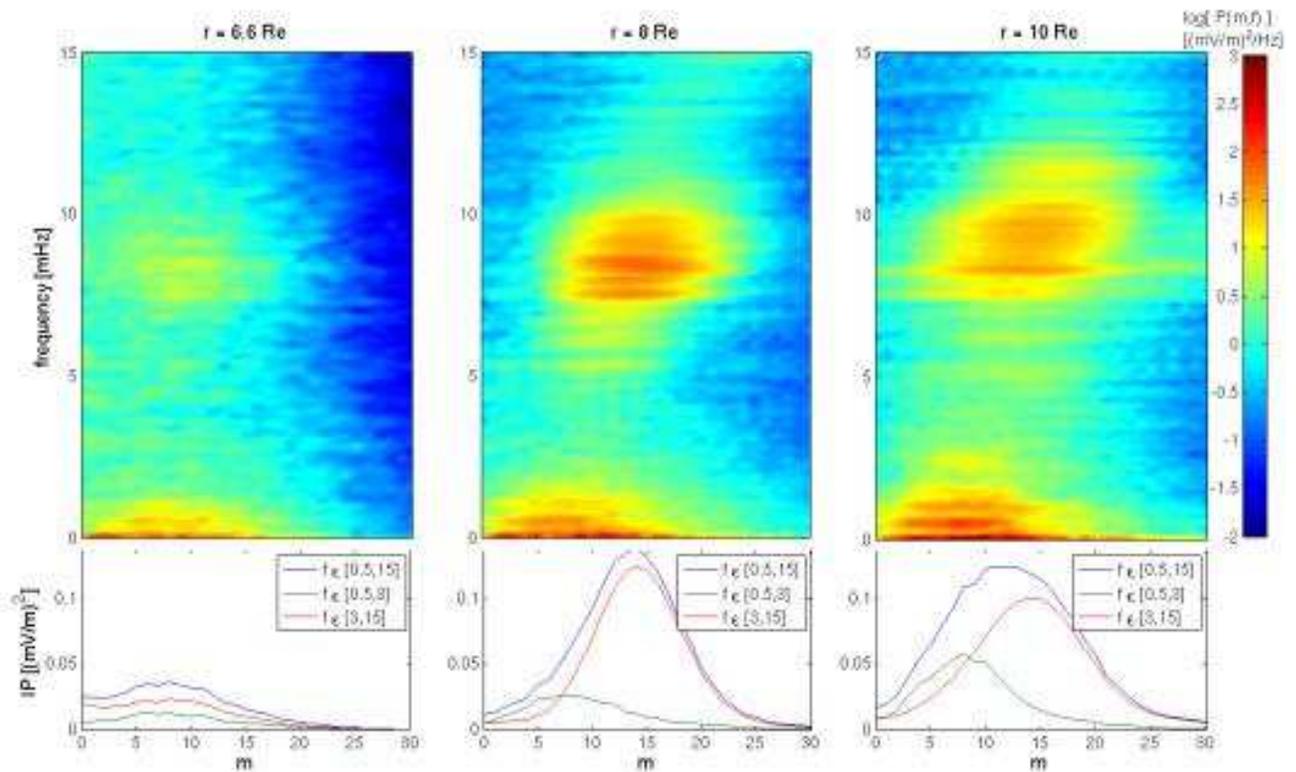}
  \caption{\label{fig:3panelaz} 
$E_{\phi}$ azimuthal mode structure in the 800 km/s simulation, along three different radii in the simulation domain: 6.6, 8, and 10 $R_{E}$.  In the color scale panels, the horizontal axis is azimuthal mode number, $m$, the vertical axis is frequency and logarithmic power spectral density is on the color scale.  The three panels beneath each of the color scale panels show integrated wave power ($IP$) over three different frequency bands, 0.5--3 mHz (green), 3--15 mHz (red), and 0.5 --15 mHz (blue), to distinguish between the MSP (green) and KH (red) populations.  Note that the MSP population has its peak wave power near $m$ $\approx$ 8 whereas the KH population has its peak wave power near $m$ $\approx$ 15 (far right panel).  The same is true for the 400 km/s and 600 km/s simulations (not shown here).  The color scale is the same in all three panels and ranges from -2 to 3.}
\end{figure*}

In \sec{2khmodes}, we compared the frequencies computed directly from the simulation results (\tbl{innerfreq} and \tbl{mpfreq}) with the peak frequencies from the power spectral density computations.  Similarly, we can compare the peaks in azimuthal mode number for the KH modes (\fig{3panelaz}) with the wave numbers computed directly from the simulation results (\tbl{innerkays} and \tbl{mpkays}).  In order to do so, we must transform the $k_{Y}$ values computed in the boundary coordinate systems into the GSM coordinate system where the azimuthal mode structure calculations were done.  Thus, we must simply decompose $k_{Y}$ into $k_{r}$ and $k_{\phi}$ = $m$/$r$.  The results of this decomposition are shown in \tbl{innerkays2} and \tbl{mpkays2} for the two KH modes.  We see that the azimuthal mode number, $m$, lies between 12 and 19 for both KH modes and all three solar wind driving speeds.  These values of $m$ are in good agreement with the peaks in power spectral density seen in \fig{3panelaz}, where we found $m$ $\approx$ 15 for both KH modes and all three solar wind driving velocities.  For a particular simulation, the values of $m$ listed in \tbl{innerkays2} and \tbl{mpkays2} show a slight difference in $m$ between the two KH modes.  This difference cannot be resolved from the power spectral density computations shown in \fig{3panelaz} due to the narrow azimuthal width separating the two KH modes.

\subsection{Criteria For KH Instability \label{sec:evalhase}}
From the simulation results, we can directly evaluate the condition for KH instability (\eq{hasegawa}) to see where it predicts the flow to be KH unstable.  As \eq{hasegawa} is only valid for a tangential discontinuity, we make no attempt to evaluate it in the simulation boundary layer.  For this calculation, we assume that there is no boundary layer and use the field values on either side of the boundary layer, outside of the boundary layer.  For example, for region 2 (the magnetosphere) fields, we use field values that are earthward of the IEBL.  Similarly, for region 1 (the magnetosheath) we use fields that are away from the magnetopause and in the magnetosheath proper.  \eq{hasegawa} cannot predict whether the inner KH mode or the magnetopause KH mode or both are excited.  It can only predict whether the field values in the magnetosheath proper and the magnetosphere proper are such that the KH instability will or will not occur.  There is only one KH mode in the incompressible, tangential discontinuity KH theory that is used to derive \eq{hasegawa}.

\begin{figure}[hl!]
  \includegraphics[scale=1.0]{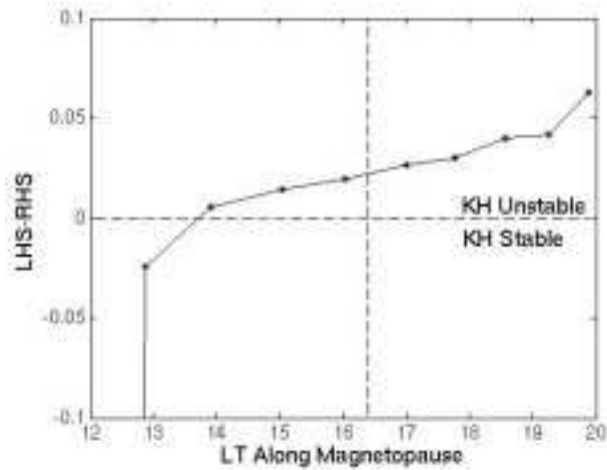}
   \caption{\label{fig:haseeval} Evaluation of the condition for KH instability (\eq{hasegawa}, {\bf k} $\mid \mid$ {\bf v}) along the equatorial plane dusk magnetopause for the 800 km/s simulation.  The solid trace shows left-hand side (LHS) minus right-hand side (RHS) from \eq{hasegawa}.  The horizontal dashed line corresponds to marginal stability.  The vertical dashed line marks the point along the dusk magnetopause where the surface waves are first seen in the simulation (1624 LT, the WAR).  Note that the condition for KH instability is first satisfied somewhere near 1400 LT.  In \sec{evalhase} we explain why the KH waves are not seen in the simulation until points near 1624 LT.}
\end{figure}

All of the field quantities in \eq{hasegawa} are specified by the simulation results.  For the wave vector {\bf k}, we use the $k_{Y}$ values listed in \tbl{innerkays} for the inner KH mode.  We choose the inner mode $k_{Y}$ values as the inner mode is predicted to be the more unstable of the two modes \citep{lee:81a}.  We evaluate this condition along the equatorial plane magnetopause, from subsolar past the dusk flank, and we assume that {\bf k} is parallel to {\bf v}.  This is a reasonable assumption given that the calculation is done in the equatorial plane and that the fastest growing mode will occur for this orientation of {\bf k} and {\bf v}.  The results of this calculation  are shown in \fig{haseeval} for a particular timestep in the 800 km/s simulation.  The horizontal axis is LT along the magnetopause and the vertical axis is the left-hand side (LHS) minus the right-hand side (RHS) in \eq{hasegawa}.  The horizontal dashed line corresponds to marginal stability.  The vertical dashed line marks the point along the magnetopause where the KH surface waves are first seen in the simulation, i.e. the WAR, as defined above.  For the 800 km/s simulation, this point is located at 1624 LT along the magnetopause.  The trace of LHS-RHS shows that the condition for KH instability is first satisfied somewhere near 1400 LT along the magnetopause.  We now address the question of why the KH waves are not seen in the simulation until points near 1624 LT on the magnetopause; \fig{haseeval} suggests they should first appear somewhere near 1400 LT.

We begin by noting that a positive value of LHS-RHS in \eq{hasegawa} is the square of the linear growth rate of the KH waves.  Thus, \fig{haseeval} shows the square of the linear growth rate of the KH waves as a function of distance along the equatorial magnetopause.  In \fig{haseeval}, we see that near 1400 LT, where the condition for KH instability is first met, the square of the growth rate is $\approx$ 0.0055, so that the growth rate is $\approx$ 0.0742 in this region.  Thus, the e-folding time in this region is $\approx$ 2$\pi$ / 0.0742 = 85 seconds.  We can now calculate the growth length in the region between 1400-1624 LT from this e-folding time and an estimation of the phase speed near 1400 LT.  A plot of the magnetosheath speed parallel to the magnetopause (not shown here) shows the value of the magnetosheath flow speed to be $\approx$ 260 km/s near 1400 LT.  Thus, the value of the KH phase velocity in this region is $\approx$ 260/2 km/s = 130 km/s \citep{walker:81a}.  These two calculations imply that the growth length in the region between 1400-1624 LT is $\approx$ 130 km/s * 85 s = 1.7 $R_E$.  Thus, the waves will travel along the magnetopause a distance of roughly 1.7 $R_E$ from 1400 LT before they can grow to a sufficient size to be resolved in the simulation.  Finally, we note that the magnetopause arc length between 1400 and 1624 LT is roughly 5.4 $R_E$.  This partially explains why the KH waves are not seen in the simulation until points near 1624 LT along the magnetopause.  The waves do not grow to a resolvable size until they travel roughly 1.7 $R_E$ along the magnetopause.  We now calculate an improved estimate of the growth length based on a more applicable KH theory in order to explain the disparity between the growth length of 1.7 $R_E$ predicted by \eq{hasegawa} and the value of 5.4 $R_E$.

The disparity between where the waves are seen in the simulations and the growth length calculation done above is probably due to the unrealistic assumptions used in deriving \eq{hasegawa}.  \eq{hasegawa} is valid for incompressible plasmas separated by a tangential discontinuity.  The LFM simulation solves the compressible MHD equations and the resolves a realistic magnetopause boundary layer.  The KH theory of \citet{walker:81a}, which solves the compressible MHD equations in the presence of a boundary layer, is a more accurate description of the KH instability at the magnetopause.  In particular, \citet{walker:81a} finds maximum (normalized) wave growth rates for $\gamma D / V_o$ in the range 0.1--0.3, where $\gamma$ is the growth rate, $D$ is half the boundary layer thickness, and $V_o$ is half the relative velocity between the two plasmas.  Near 1400 LT, where \eq{hasegawa} first predicts the flow to be KH unstable, the value of $\gamma D / V_o$ is roughly 0.9, using the $\gamma$ value near 1400 LT (0.0742), and the simulation values near 1400 LT for $D$ (3121/2 km) and $V_o$ (130 km/s).  Thus, \eq{hasegawa} predicts a normalized growth rate that is much larger than what is reported in \citet{walker:81a}.  Assuming a normalized growth rate of $\gamma D / V_o$ = 0.25 and using the LFM simulation results near 1400 LT for $D$ and $V_o$, we calculate a growth length of 6.2 $R_E$.  This calculation of the growth length is in better agreement with the distance between where \eq{hasegawa} first predicts the flow to be KH unstable and where the waves are first seen in the simulation, a distance of roughly 5.4 $R_E$.  Similar results hold for the 400 km/s and 600 km/s simulations (not shown here).  

At this point, it should be clear that the surface waves seen near the dawn and dusk flanks in the three simulations are indeed Kelvin-Helmholtz waves.  The simulation surface wave characteristics are consistent with the theoretical and observational KH surface wave results.  The simulated waves have the proper frequencies, wavelengths, phase velocities, propagation directions and they have the large vortical structures associated with them.  Furthermore, we see maximum wave growth for values of $kd$ consistent with theoretical predictions.  We also find that the theoretical results predict the magnetopause boundary to be KH unstable and the theoretical growth rate of the waves is consistent with where the waves are first seen in the simulations.  Again, we emphasize the fact that the solar wind dynamic pressure is constant in our simulations.  Thus, the surface waves cannot be attributed to fluctuations in the solar wind dynamic pressure, a claim that is often used to discount observational evidence of KH generated surface waves \citep[e.g.][]{song:88a}.

As an aside, we note that the KH instability has been invoked to explain surface waves and vortical structures seen in global MHD simulations driven by real solar wind conditions.  \citet{slinker:03a} compared LFM simulation results with Geotail observations of magnetopause crossings.  The LFM simulation reproduced the surface waves observed by Geotail and the authors noted that the likely source of the oscillations was the KH instability.  Similarly, \citet{collado-vega:07a} simulated 9 hours of a high speed solar wind stream that was seen at L1 between 29 March to 5 April 2002, using the LFM simulation.  The authors reported large vortical structures near the magnetopause boundary and attributed these vortices to the KH instability.  In both of these studies, the authors suggested the the KH instability was responsible for the surface waves and vortical structures but offered no conclusive evidence that the KH instability was indeed the source.

Finally, we note that all of the KH theory discussed in this paper is linear MHD wave theory.  Thus, once the KH waves have developed into their nonlinear stage, the linear wave theory is no longer applicable.  The formation of the large vortical structures in the simulation is strong evidence that we have reached the nonlinear stage \citep{miura:84a,wu:86a}.  Thus, applying the linear theory at points along the magnetopause boundary where the waves have reached their nonlinear stage is invalid.  

\section{Summary and Conclusions \label{sec:conclusions}}

In this paper, global, three-dimensional MHD simulations of the solar wind/\-mag\-neto\-sphere interaction were used to study ULF pulsations in the inner magnetosphere.  The MHD simulations were driven with idealized, constant solar wind input parameters.  These parameters were chosen to study the effect of changing only the solar wind driving velocity, while holding the other solar wind input parameters constant.  Driving the simulations with constant solar wind parameters ensured that any discrete ULF pulsations in the simulation magnetosphere were not driven by fluctuations in the solar wind.  The simulation results revealed ULF surface waves near the dawn and dusk flank magnetopause.  These surface waves were shown to be driven by the Kelvin-Helmholtz instability and not dynamic pressure fluctuations in the solar wind.

A closer examination of the surface waves revealed that two KH modes were seen near the dawn and dusk flank magnetopause.  These two KH modes were identified as the inner KH mode and the magnetopause KH mode, as described in \citet{lee:81a,kivelson:84a}.  The magnetopause KH mode was found to propagate tailward along the magnetopause boundary whereas the inner KH mode was found to propagate tailward along the inner edge of the boundary layer (IEBL).  These two KH modes were found to have different phase velocities and different wavelengths but oscillated at the same frequency.  We presented a scientific visualization that showed the coupled oscillation of the two KH modes and a coupled oscillation of the low-latitude boundary layer.  The scientific visualization also revealed large vortical structures associated with the inner KH mode.  These vortical structures were centered on the IEBL and propagated tailward along the IEBL, growing in size as they moved downtail.  Both KH modes were found to occur for $kd$ = 0.5--1.0 where $k$ is the wave number and $d$ is the boundary layer thickness.  This fact was used to explain the monochromatic nature of the KH waves.  The frequency of the KH waves was found to depend on the solar wind driving velocity, with larger driving velocities generating KH waves with higher frequencies.  The azimuthal mode number, $m$, of the KH waves was found to be between 15--20 and did not change significantly with solar wind driving speed.  The relativistic first adiabatic invariant, $M$, was computed from the $m$ and $f$ values of these KH waves.  We found that the KH waves could effectively interact with equatorial plane radiation belt electrons of a few hundred keV, near the dusk meridian.

\begin{table}[h!]
\begin{tabular}{|l|c|c|c|c|}	\hline
& \{$m$,$f$\} & $L_{min}$ & $B(L_{min})$ & $M$  \\
&  &  & (nT) & (MeV/G) \\ \hline
400 Run & \{15,5\} & 10.5 & 31 & 377 \\ \hline
600 Run & \{15,8\} & 8.6 & 52 & 469 \\ \hline
800 Run & \{15,10\} & 7.5 & 79 & 513 \\ \hline 
\end{tabular}
\caption{Relativistic first adiabatic invariant $M$ values computed for the given \{$m$,$f$\} pair, for the three simulations in this study.  These $M$ values determine the electron populations that the KH waves could interact with.  We also show the $L$ and $B$ values along the dusk meridian where we assume the interaction occurs.  These values are for the most inward radial penetration of the inner KH mode.}
\label{tbl:innerMvals}
\end{table}

\begin{table}[h!]
\begin{tabular}{|l|c|c|c|c|}	\hline
& \{$m$,$f$\} & $L_{max}$ & $B(L_{max})$ & $M$  \\
&  &  & (nT) & (MeV/G) \\ \hline
400 Run & \{15,5\} & 13.5 & 15 & 601 \\ \hline
600 Run & \{15,8\} & 11.6 & 16 & 738 \\ \hline
800 Run & \{15,10\} & 10.5 & 24 & 835 \\ \hline 
\end{tabular}
\caption{Same as \tbl{innerMvals} except the $M$ values are computed near the magnetopause}
\label{tbl:outerMvals}
\end{table}

\begin{table}[h!]
\begin{tabular}{|l|c|c|c|}	\hline
& {\bf 400 km/s} & {\bf 600 km/s} & {\bf 800 km/s} \\ \hline
$v_{phase}$ (km/s) & 140 & 160 & 225 \\ \hline
$\lambda_{Y}$ ($R_{E}$) & 4.2 & 3.3 & 3.3 \\ \hline
$f$ (mHz) & 5.2 & 7.6 & 10.7 \\ \hline 
\end{tabular}
\caption{Inner KH mode equatorial plane phase velocities and wavelengths, computed directly from the simulation results, and the resulting wave frequencies.}
\label{tbl:innerfreq}
\end{table}

\begin{table}[h!]
\begin{tabular}{|l|c|c|c|}	\hline
& {\bf 400 km/s} & {\bf 600 km/s} & {\bf 800 km/s} \\ \hline
$v_{phase}$ (km/s) & 180 & 225 & 375 \\ \hline
$\lambda_{Y}$ ($R_{E}$) & 5.2 & 4.3 & 5.2 \\ \hline
$f$ (mHz) & 5.4 & 8.2 & 11.3 \\ \hline 
\end{tabular}
\caption{Same as \tbl{innerfreq} for the magnetopause KH mode.}
\label{tbl:mpfreq}
\end{table}

\begin{table}[h!]
\begin{tabular}{|l|c|c|c|}	\hline
& {\bf 400 km/s} & {\bf 600 km/s} & {\bf 800 km/s} \\ \hline
$d$ ($R_{E}$) & 0.53 & 0.48 & 0.47 \\ \hline
$k_{Y}$ (1/$R_{E}$) & 1.50 & 1.90 & 1.90 \\ \hline
$kd$ & 0.80 & 0.91 & 0.90 \\ \hline
\end{tabular}
\caption{Simulation boundary layer thickness, $d$, the Y component of the wave vector {\bf k} in the boundary coordinate system, and the product $kd$, for the inner KH mode in the three simulations (under the assumption $k$ $\approx$ $k_{Y}$; see \fig{kz}).  Note the values of $kd$ in the range 0.5--1.0.}
\label{tbl:innerkays}
\end{table}

\begin{table}[h!]
\begin{tabular}{|l|c|c|c|}	\hline
& {\bf 400 km/s} & {\bf 600 km/s} & {\bf 800 km/s} \\ \hline
$d$ ($R_{E}$) & 0.53 & 0.48 & 0.47 \\ \hline
$k_{Y}$ (1/$R_{E}$) & 1.21 & 1.46 & 1.21 \\ \hline
$kd$ & 0.64 & 0.70 & 0.57 \\ \hline
\end{tabular}
\caption{Same as \tbl{innerkays} for the magnetopause KH mode.}
\label{tbl:mpkays}
\end{table}

\begin{table}[h!]
\begin{tabular}{|l|c|c|c|}	\hline
& {\bf 400 km/s} & {\bf 600 km/s} & {\bf 800 km/s} \\ \hline
$k_{r}$ (1/$R_{E}$) & 0.68 & 0.86 & 1.04 \\ \hline
$k_{\phi}$ (1/$R_{E}$) / $m$ & 1.33 / 18 & 1.69 / 19 & 1.6 / 16 \\ \hline 
\end{tabular}
\caption{The equatorial plane components of the wave vector {\bf k} in the GSM coordinate system for the inner KH mode in the three simulations. Note that the azimuthal mode number, $m$, lies between 12 and 19 for all three solar wind driving speeds, in good agreement with the $m$ peaks in \fig{3panelaz}.}
\label{tbl:innerkays2}
\end{table}

\begin{table}[h!]
\begin{tabular}{|l|c|c|c|}	\hline
& {\bf 400 km/s} & {\bf 600 km/s} & {\bf 800 km/s} \\ \hline
$k_{r}$ (1/$R_{E}$) & 0.66 & 0.79 & 0.70 \\ \hline
$k_{\phi}$ (1/$R_{E}$) / $m$ & 1.02 / 16 & 1.22 / 17 & 0.99 / 12 \\ \hline 
\end{tabular}
\caption{Same as \tbl{innerkays2} for the magnetopause KH mode.}
\label{tbl:mpkays2}
\end{table}

\acknowledgments
This material is based upon work supported by the National Aeronautics and Space Administration under Grant Nos. NNG05GK04G and NNX07AG17G and by the Center for Integrated Space Weather Modeling which is funded by the Science and Technology Centers program of the National Science Foundation under Agreement number ATM-0120950.  The National Center for Atmospheric Research is sponsored by the National Science Foundation.  The authors are grateful for thoughtful discussions with Drs. W. Lotko, J. G. Lyon, I. R. Mann, and R. L. McPherron.


\clearpage

\end{document}